\documentclass[twocolumn,showpacs,preprintnumbers,amsmath,amssymb]{revtex4}

\usepackage[dvips]{color}
\definecolor{Red}{rgb}{1.00, 0.00, 0.00}

\usepackage{graphicx}
\usepackage{dcolumn}
\usepackage{bm}

\newcommand{\ep}{\varepsilon}

\newcommand{\eps}{\epsilon}

\newcommand{\la}{\lambda}

\newcommand{\si}{\sigma}

\renewcommand{\th}{\theta}   

\newcommand{\om}{\omega}

\newcommand{\p}{\partial}
 
\newcommand{\dsp}{\displaystyle}
\newcommand\eqn[1]{(\ref{#1})}      
\newcommand{\beq}{\begin{equation}}
\newcommand{\eeq}{\end{equation}}
\newcommand{\ba}{\begin{array}}
\newcommand{\ea}{\end{array}}
\newcommand{\bea}{\begin{eqnarray}}
\newcommand{\eea}{\end{eqnarray}}
\newcommand{\bi}{\begin{itemize}}  
\newcommand{\ei}{\end{itemize}}
\newcommand{\ben}{\begin{enumerate}} 
\newcommand{\een}{\end{enumerate}}

\newcommand{\feyn}[1]{
  \setbox0=\hbox{\ensuremath{#1}}
  \hbox to\wd0{\hbox to0pt{\hbox to\wd0{\hss/\hss}\hss}\box0}}

\newcommand{\Qtilde}{{\tilde Q}}
\newcommand{\almix}{\alpha}

\begin{document}

\preprint{}

\title{Illuminating interfaces between phases of a U(1)$\times$U(1) gauge theory
}

\author{Mark Alford}\author{Gerald Good}%
\affiliation{Physics Department, Washington University,
St.~Louis, MO~63130, USA}

\date{Feb 2004}

\begin{abstract}
We study reflection and transmission of light
at the interface between
different phases of a $U(1)\otimes U(1)$ gauge theory.
On each side of the interface, one can choose a basis so that
one generator is free (allowing propagation
of light), and the orthogonal one
may be free, Higgsed, or confined. However, the basis on one side
will in general be rotated relative to the 
basis on the other by some angle $\almix$.
We calculate reflection and transmission coefficients for
both polarizations of light and all 8 types of boundary, for arbitrary $\almix$.
We find that an observer measuring the behavior of light beams
at the boundary would be able to distinguish 4 different types
of boundary, and we show how the remaining ambiguity arises
from the principle of complementarity
(indistinguishability of confined and Higgs phases) which
leaves observables invariant under a global
electric/magnetic duality transformation.
We also explain the seemingly paradoxical behavior
of Higgs/Higgs and confined/confined boundaries, and clarify some
previous arguments that confinement must involve magnetic monopole
condensation.
\end{abstract}


\maketitle

\section{Introduction}
\label{sec:intro}

In this paper we study boundaries between phases in which
different linear combinations of gauge generators are free.
Mixing of gauge generators is familiar from the
standard model of particle physics, and the possibility of
creating neighboring domains in which different 
linear combinations of gauge generators are free
is now receiving serious attention. To set the stage
for this work we first briefly review a concrete example.

In the standard model, the propagating $U(1)$
gauge boson (the photon) is associated with a particular Abelian
$U(1)_Q$ subgroup of the full standard model gauge group. This
subgroup emerged unbroken from the electroweak Higgs symmetry breaking
$SU(2)\otimes U(1)_Y \to U(1)_Q$ at the TeV scale, and is generated by
some linear combination of the ``$W_3$'' generator of the $SU(2)$ weak
interaction and the ``$Y$'' generator of the $U(1)$ hypercharge
interaction. 

We now know that in quark matter (which may well occupy
macroscopic regions of space, inside neutron stars)
the gauge group for the propagating $U(1)$ gauge boson will
be rotated into 
a different direction by a further layer of symmetry breaking
at the MeV scale.
At sufficiently high density, quark matter will
develop a condensate
of quark Cooper pairs that plays the role of a Higgs field
\cite{oldcolorSC,newcolorSC}. (For reviews of this phenomenon
of ``color superconductivity'' see Ref.~\cite{Reviews}).
In the real world, quark matter is expected
to contain the three lightest flavors, and in this case
the condensate forms a ``color-flavor-locked''
(CFL) phase \cite{Alford:1998mk}, in which a linear
combination of the photon and one of the gluons remains massless,
while the orthogonal linear combination and the remainder of the
gluons become massive by the Higgs mechanism. 
The gauge symmetry breaking
is $SU(3)_{\rm color}\otimes U(1)_Q \to U(1)_\Qtilde$.
Thus a ``rotated" electromagnetism is present in
the CFL color superconducting phase of quark matter.
This raises the interesting possibility of having an interface between
a vacuum region in which the propagating gauge boson is the usual $Q$-photon,
and a quark matter region in which it is a different particle, 
the $\Qtilde$-photon, which is a mixture of the photon and a gluon.
What will happen to electromagnetic fields,
including light beams, that encounter such an interface?

This question is not a completely theoretical one: it has often been
speculated that three-flavor quark matter could be absolutely stable,
so a quark matter star (``strange star'') could have a surface
at which the CFL phase meets the vacuum. Since the CFL phase
is a transparent insulator \cite{CFLneutral} we could in principle
literally see into the core of such a star.

The $U(1)\otimes U(1)$ gauge system arises in various
other physical contexts.
Electroweak symmetry
breaking can be simplified to a $U(1)\otimes U(1)$ system by
focussing on the hypercharge and $W_3$ bosons, which mix to form
the photon and $Z^0$. The $U(1)\otimes U(1)$ gauge system also
arises in extensions of the standard model, where an extra $U(1)$
gauge symmetry with a corresponding
$Z'$ gauge boson is added. Natural contexts for this include
Grand Unified
Theories with gauge groups such as $SO(10)$ and $E_6$, and some string
models \cite{Hewett,Babu:1997st}. 

In this paper we study the light reflection and transmission
properties of a boundary
between phases in a $U(1)\otimes U(1)$ gauge theory.
There have been previous
studies of the behavior of magnetic fields \cite{Alford:1999pb}
and light beams \cite{manraj} in the specific case of the
interface between the vacuum and CFL quark matter.
However, we consider the most general
realization of the gauge symmetries that supports propagating gauge bosons.
On one side of the boundary both $U(1)$ gauge symmetries 
may be free, or some linear
combination may be Higgsed or confined. On the other side, both
$U(1)$ gauge symmetries may be free, or a {\em different} linear
combination may be Higgsed or confined, where
the difference is parameterized by a ``mismatch angle'' $\almix$.
We calculate the nature and intensity of the reflected and transmitted
gauge bosons in each case.

In section \ref{sec:bc} we introduce the $U(1)\otimes U(1)$ model
and show how Higgsing or confinement of
a gauge field can be implemented by appropriate boundary
conditions at the interface. Section \ref{sec:results} describes
the calculation of the reflection and transmission coefficients for 
the various types of boundary.
In section \ref{sec:discussion} we summarize our results.
We then discuss how they compare with previous calculations,
explain some mysterious features, and analyze their
compatibility with expectations based on the complementarity
principle. Appendices \ref{app:frequency} and \ref{app:fields}
analyze subleties of the low-frequency limit and a detailed
example of complementarity.

\section{Confinement and Higgsing via Boundary Conditions
in the $U(1)\otimes U(1)$ model}
\label{sec:bc}

We place the interface at the $z=0$ plane. 
On the $z>0$ side of the interface we work in the  $U(1)_Q\otimes U(1)_{T_8}$
basis. $U(1)_Q$ is free, so that photons can propagate, and
$U(1)_{T_8}$ may be confined, Higgsed, or free.
On the $z<0$ side of the interface we work in the $U(1)_\Qtilde\otimes U(1)_X$
basis. $U(1)_\Qtilde$ is free, so that $\Qtilde$-photons can propagate,
and $U(1)_X$ may be confined, Higgsed, or free
\footnote{These names for the generators are taken over from
earlier treatments of the interface between
a vacuum and CFL quark matter \cite{Alford:1999pb,manraj}, 
where one only considers the
electromagnetic generator $Q$ and the color generator $T_8$ with
which it mixes.  In the CFL quark matter, a quark-quark condensate
acts as a Higgs field, breaking a linear combination $X$
and leaving the orthogonal linear combination $\Qtilde$ unbroken.}%
.
The $\Qtilde$-photon 
\beq
  A_{\mu}^{\tilde{Q}} = \cos\almix A_{\mu} + \sin\almix G_{\mu}^8
\eeq
remains free, while the orthogonal ``$X$'' gauge boson
\beq
   A_{\mu}^X  = -\sin\almix A_{\mu} + \cos\almix G_{\mu}^8
\eeq
may be free, Higgsed, or confined.
In the case of CFL matter, electromagnetism ($Q$)
is much more weakly coupled than the strong 
interaction ($T_8$) at the relevant energy scale, so
the mixing angle $\almix$ (analogous to the Weinberg angle in the standard
model) is small, and the $\Qtilde$ photon is mostly the 
ordinary $Q$-photon, with a small admixture of the $T_8$ gluon.
However, in our general treatment, we will keep $\almix$ 
as an arbitrary parameter.

\begin{figure}[htb]
\includegraphics[width=0.5\textwidth]{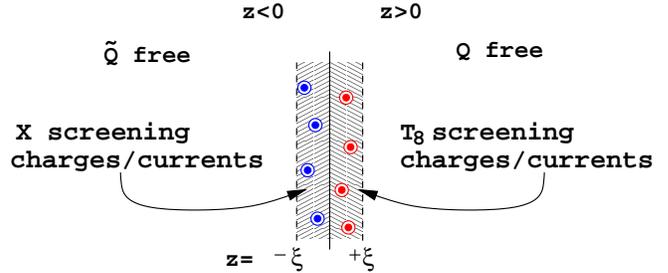}
\caption{The phase boundary that we study. In the $z>0$ region,
the $Q$ gauge boson is free, and the orthogonal $T_8$ gauge boson
may be free, or it may be Higgsed or confined. In the $z<0$ region,
the $\Qtilde$ gauge boson is free, and the orthogonal $X$ gauge boson
may be free, or it may be Higgsed or confined. Higgsing and confinement
are implemented by currents or charges in the boundary region
of thickness $\xi$. The condensates that cause Higgsing/confinement
are assumed to change over a much shorter distance.
}
\label{fig:boundary}
\end{figure}

We study the behavior of $Q$-photons coming in from $z=+\infty$,
and reflecting off or transmitting through the interface.
As in Ref.~\cite{Alford:1999pb}, we use free Maxwell
equations to describe all the gauge fields,
with confinement and Higgsing implemented
via boundary conditions at the interface
(Fig.~\ref{fig:boundary}), as we now describe.

At the boundary of a free phase, there are no limitations on the electric and
magnetic fields of both $U(1)$ generators: both types
of gauge boson can propagate, and there are no charges or
currents present.

At the boundary of a Higgsed phase, there is a layer of thickness
$\xi$ in which 
there are electric charges and super-currents associated with
the Higgsed generator. This corresponds to the real physics
of a Higgs phase, in which a condensate of a charged field
supplies mobile electric charges that screen out electric flux
and repel magnetic flux (the Meissner effect).
For a confined phase, there is a boundary layer of thickness
$\xi$ in which 
there are magnetic charges and super-currents associated with
the confined generator.  This corresponds to the dual
superconductor picture of confinement \cite{thftmand},
in which there are mobile magnetic charges that screen
out magnetic flux and repel electric flux.

Note that we assume the ``sharp interface'' scenario of
Ref.~\cite{Alford:1999pb}, in which the wavelength $\lambda$ 
of the light shining on the
boundary is much larger than the penetration
depth $\xi$ for the gauge fields.
This assumption seems straightforward but actually
under some circumstances there are subtle order-of-limits issues.
We will discuss them in section \ref{sec:discussion}
when we address the paradoxical
nature of the $\almix\to 0$ limit for certain interfaces.

To proceed, we write all fields as two-component objects
in the two-dimensional
space of gauge symmetry generators spanned by $Q$ and $T_8$.
The $(\Qtilde,X)$ basis is rotated by the angle $\almix$:
\beq
\label{basis}
\ba{rcl@{\quad}rcl}
 Q &=& \left(\ba{@{}c@{}}1 \\ 0 \ea \right), &
T_8 &=& \left(\ba{@{}c@{}}0 \\ 1 \ea \right), \\[2ex]
\Qtilde &=&  \left(\ba{@{}c@{}}\cos\almix \\ \sin\almix \ea \right), &
X &=&  \left(\ba{@{}c@{}} -\sin\almix \\ \phantom{-}\cos\almix \ea \right).
\ea
\eeq
so a general magnetic field takes the form
\beq
   \vec{B} = \begin{pmatrix} {\vec{B}}^Q \\ {\vec{B}}^{T_8} \end{pmatrix}
           = \begin{pmatrix} 
             \cos\almix {\vec{B}}^{\tilde{Q}} - \sin\almix {\vec{B}}^X \\
             \sin\almix {\vec{B}}^{\tilde{Q}} + \cos\almix {\vec{B}}^X 
                            \end{pmatrix}
\eeq
and similarly for $\vec{E}$. The generalized Maxwell equations are
\bea
   \label{eqn:maxwell}
   \nabla \cdot \vec{D} = \rho&,& 
   \nabla \times \vec{E} = \vec{J}_M - \frac{\p \vec{B}}{\p t},\\
   \nabla \cdot \vec{B} = \rho_M&,& \nabla \times \vec{H} = \vec{J} + \frac{\p \vec{D}}{\p t}
\eea
where $\rho_M$ and $J_M$ are magnetic charge and current densities,  
and we assume the usual linear relationship between 
$\vec{E}$ and $\vec{D}$,  and between $\vec{B}$ and $\vec{H}$,
\beq
\ba{rcl@{\qquad}rcl}
  \multicolumn{3}{l}{\mbox{$Q$-photons:}} & 
    \multicolumn{3}{l}{\mbox{$\Qtilde$-photons:}}\\[1ex]
  \vec{D}^Q &=& \ep \vec{E}^Q, 
    & \vec{D}^\Qtilde &=& \tilde\ep \vec{E}^\Qtilde, \\[1ex]
  \vec{H}^Q &=& \dsp\frac{1}{\mu} \vec{B}^Q, 
    & \vec{H}^\Qtilde &=& \dsp\frac{1}{\tilde\mu} \vec{B}^\Qtilde.
\ea
\eeq
We assume that the wavelength of the gauge bosons incident on the surface
is much greater than the penetration depth $\xi$
so we can integrate the Maxwell equations
over $-\xi < z < \xi$, and obtain
boundary conditions that relate the fields at $z=-\xi$ to
those at $z=+\xi$ (Ref.~\cite{Jackson}, sect.~I.5).
For the fields with divergence equations ($D$ and $B$) the
boundary conditions relate the components perpendicular to
the surface; for the fields with curl equations ($E$ and $H$)
boundary conditions relate the components parallel to
the surface.
\newcommand{\Qvec}{ \begin{pmatrix}1\\0\end{pmatrix} }
\newcommand{\Tvec}{ \begin{pmatrix}0\\1\end{pmatrix} }
\newcommand{\Qtvec}{\begin{pmatrix}\cos\almix \\ \sin\almix \end{pmatrix} }
\newcommand{\Xvec}{ 
  \begin{pmatrix}-\sin\almix \\ \phantom{-}\cos\almix\end{pmatrix} }
\begin{widetext}
\bea
   D^Q_{\perp}(\xi )  \Qvec  + D^{T_8}_{\perp}(\xi ) \Tvec
 - D^{\tilde{Q}}_{\perp}(-\xi ) \Qtvec  - D^X_{\perp}(-\xi )   \Xvec
 &=& \si^{T_8} \Tvec  + \si^X \Xvec \label{bc:D} \\
   E^Q_{\parallel}(\xi )  \Qvec     + E^{T_8}_{\parallel}(\xi ) \Tvec
 - E^{\tilde{Q}}_{\parallel}(-\xi ) \Qtvec  - E^X_{\parallel}(-\xi ) \Xvec
 &=& K^{T_8}_M \Tvec + K^X_M \Xvec \label{bc:E} \\
   B^Q_{\perp}(\xi ) \Qvec  + B^{T_8}_{\perp}(\xi ) \Tvec
 - B^{\tilde{Q}}_{\perp}(-\xi ) \Qtvec  - B^X_{\perp}(-\xi ) \Xvec
 &=& \si^{T_8}_M \Tvec    + \si^X_M \Xvec \label{bc:B} \\
   H^Q_{\parallel}(\xi ) \Qvec  + H^{T_8}_{\parallel}(\xi ) \Tvec
 - H^{\tilde{Q}}_{\parallel}(-\xi ) \Qtvec  - H^X_{\parallel}(-\xi ) \Xvec
 &=& K^{T_8} \Tvec  + K^X \Xvec \label{bc:H}
\eea
\end{widetext}
The $\si$'s and $K$'s are the effective surface charge and current densities,
and their presence varies depending on the physical situation being
addressed. For Higgsed generators there are electric surface
current and charge densities, for confined generators there are magnetic
surface current and charge densities, and for free generators there are no
surface current or charge densities.

\begin{figure*}[htb]
   \includegraphics[width=0.35\textwidth]{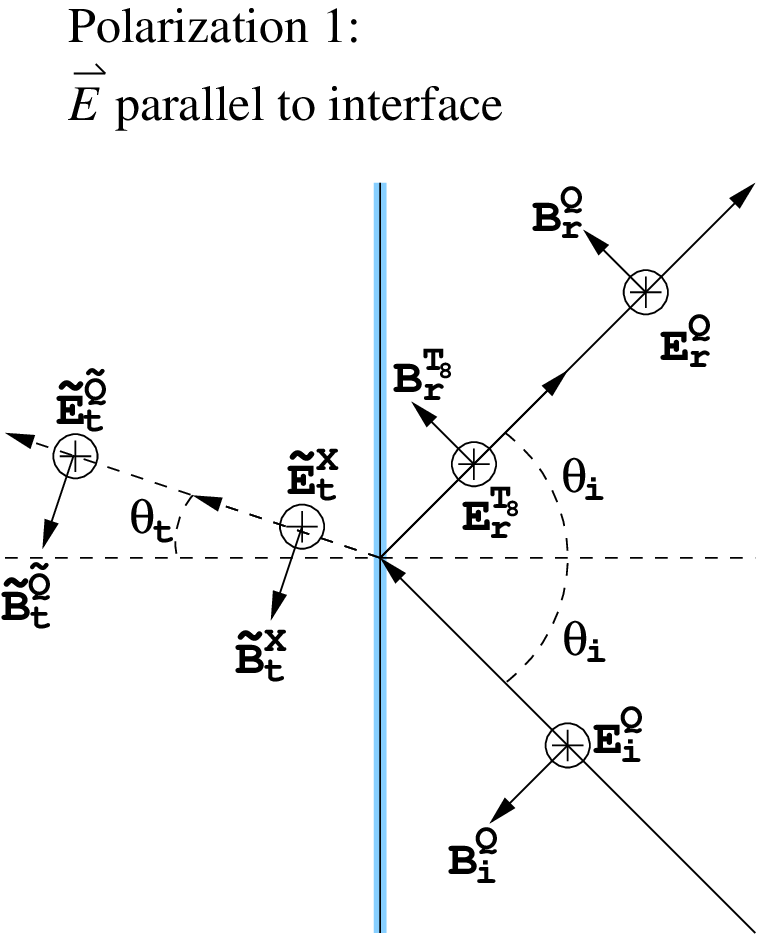}
   \hspace{3em}
   \includegraphics[width=0.35\textwidth]{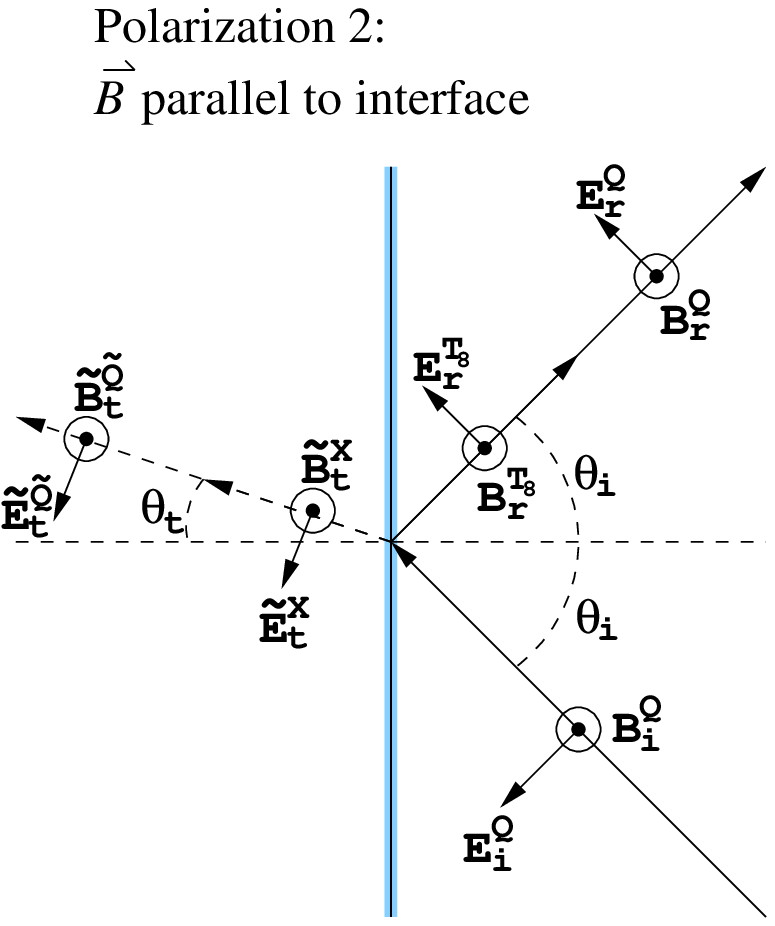}
   \caption{Polarizations of the incident photon beam}
   \label{fig:pol}
\end{figure*}

\section{Reflection and Transmission at the Interface}
\label{sec:results}



In our analysis we treated both possible polarizations, as illustrated 
in Fig.~\ref{fig:pol}. Without loss of generality,
we assume that the waves
incident from $z=+\infty$  will be 
purely $Q$ gauge bosons. For free phases, two different types of gauge 
boson may be
reflected and/or transmitted. It is also assumed that 
in a phase where both
types of gauge boson are massless, the index of refraction
(and hence the $\ep$ and $\mu$) is the same for both.

In addition, the usual rules of optics apply, since they are purely
kinematic in nature \cite{Jackson}. Therefore the angle of 
reflection equals the angle of incidence, and Snell's Law applies 
to the transmitted waves. 

To find the transmission and reflection coefficients, we
applied the boundary conditions of section \ref{sec:bc} to the kinematic
situations shown in Fig.~\ref{fig:pol}. 
Tables \ref{tab:RTcoeff-pol1} and \ref{tab:RTcoeff-pol2} show the
results
of these calculations for the eight non-trivial phase combinations. 
(Some intermediate results, the
transmission/reflection {\em amplitudes}, are
shown and discussed in appendix \ref{app:fields}).

The shorthand parameters used throughout the calculations are 
defined as follows:
\bea
\label{angledefs}
   r &\equiv& \frac{\mu}{\tilde{\mu}} \frac{\tilde{n}}{n}
     = \sqrt{\frac{\tilde{\ep} \mu}{\ep \tilde{\mu}}},\\
   c_i &\equiv& \cos \th_i, s_i \equiv \sin \th_i, \\
   c_t &\equiv& \cos \th_t, s_t \equiv \sin \th_t.
\eea
We can eliminate $\cos \th_t$ from the amplitudes by making use of
Snell's Law, 
\bea
   n \sin \th_i &=& \tilde{n} \sin \th_t \\
   \to \cos \th_t &=& \sqrt{1 - \frac{n^2}{\tilde{n}^2} \sin^2\! \th_i}.
\eea
Reflection and transmission coefficients are defined by 
\bea
\label{RT1}
   R &=& \frac{I_r}{I_i} \equiv \frac{\vec{S_r}_{\perp}}{\vec{S_i}_{\perp}} \\
   T &=& \frac{I_t}{I_i} \equiv \frac{\vec{S_t}_{\perp}}{\vec{S_i}_{\perp}} 
\eea
where $I_i, I_r, I_t$ refer to the incident, reflected, and transmitted
intensities, respectively, and $\vec{S_i}, \vec{S_r}, \vec{S_t}$ refer to
the incident, reflected, and transmitted Poynting vectors; specifically,
\beq
\label{RT2}
\ba{rclcl}
   R^Q &=& \dsp\frac{\hat{z} \cdot \vec{E}^Q_r \times \vec{H}^Q_r}
                {\hat{z} \cdot \vec{E}^Q_i \times \vec{H}^Q_i}
       &=& \dsp\frac{ c_r {{\cal{E}}^Q_r}^2}
                { c_i {{\cal{E}}^Q_i}^2}, \\[2ex]
   R^{T_8} &=& \dsp\frac{\hat{z} \cdot \vec{E}^{T_8}_r \times \vec{H}^{T_8}_r}
                {\hat{z} \cdot \vec{E}^Q_i \times \vec{H}^Q_i}
       &=& \dsp\frac{ c_r {{\cal{E}}^{T_8}_r}^2}
                { c_i {{\cal{E}}^{T_8}_i}^2}, \\[2ex]
   T^{\tilde{Q}} &=& 
    \dsp\frac{\hat{z} \cdot \vec{E}^{\tilde{Q}}_t \times \vec{H}^{\tilde{Q}}_t}
                {\hat{z} \cdot \vec{E}^Q_i \times \vec{H}^Q_i} 
       &=& \dsp \sqrt{\frac{\tilde{\eps}\mu}{\eps\tilde{\mu}}}
\frac{ c_t { { \cal{E} }^{\tilde{Q}}_t }^2}
                { c_i {{\cal{E}}^Q_i}^2}. \\[2ex]
\ea
\eeq
For clarity, we will illustrate how the calculations leading to tables
\ref{tab:RTcoeff-pol1} and \ref{tab:RTcoeff-pol2} are done by looking at
two of the cases in detail. The amplitudes ${\cal E}$ give the
electric fields associated with the incident, reflected, and transmitted
photons,
\beq
\ba{rcl}
\vec{E}^Q_i &=& {\cal{E}}^Q_i \vec n_i 
  \exp(i(\vec{k}_i \cdot \vec{x}-\om t))\ ,\\
\vec{E}^{(Q,T_8)}_r &=& {\cal{E}}^{(Q, T_8)}_r \vec n_r
  \exp(i(\vec{k}_r \cdot \vec{x}-\om t))\ , \\
\vec{E}^{\tilde{Q}}_t &=& {\cal{E}}^{\tilde{Q}}_t \vec{n}_t 
  \exp(i(\vec{k}_t \cdot \vec{x}-\om t))\ ,
\ea
\eeq
where $\vec{n}$ is the unit polarization vector for each wave.

\subsection{ $T_8$ Free, $X$ Confined}
For this combination of phases, the boundary condition equations
\eqn{bc:D}, \eqn{bc:E}, \eqn{bc:B}, \eqn{bc:H} become
\begin{widetext}
\bea
   D^Q_{\perp}(\xi )  \Qvec  + D^{T_8}_{\perp}(\xi ) \Tvec
 - D^{\tilde{Q}}_{\perp}(-\xi ) \Qtvec &=& 0 \\
   E^Q_{\parallel}(\xi )  \Qvec     + E^{T_8}_{\parallel}(\xi ) \Tvec
 - E^{\tilde{Q}}_{\parallel}(-\xi ) \Qtvec  &=& K^X_M \Xvec \\
   B^Q_{\perp}(\xi ) \Qvec  + B^{T_8}_{\perp}(\xi ) \Tvec
 - B^{\tilde{Q}}_{\perp}(-\xi ) \Qtvec &=& \si^X_M \Xvec \\
   H^Q_{\parallel}(\xi ) \Qvec  + H^{T_8}_{\parallel}(\xi ) \Tvec
 - H^{\tilde{Q}}_{\parallel}(-\xi ) \Qtvec &=& 0 
\eea
\end{widetext}
Dotting the equations with either 
${\begin{pmatrix}\cos\almix \\ -\sin\almix \end{pmatrix} }$, $\Qvec$, or
$\Tvec$ as appropriate yields equations that do not depend on the charge or
current densities. In this case, we obtain
\bea
   H^Q_{\parallel}(\xi ) &=& \cos\almix H^{\tilde{Q}}_{\parallel}(-\xi ) 
      \label{ex1:one}\\
   H^{T_8}_{\parallel}(\xi ) &=& \sin\almix H^{\tilde{Q}}_{\parallel}(-\xi ) 
      \label{ex1:two}\\ 
   \cos\almix E^Q_{\parallel}(\xi ) + \sin\almix E^{T_8}_{\parallel}(\xi )
   &=& E^{\tilde{Q}}_{\parallel}(-\xi ) \label{ex1:three} \\
   \cos\almix B^Q_{\perp}(\xi ) + \sin\almix B^{T_8}_{\perp}(\xi ) 
   &=& B^{\tilde{Q}}_{\perp}(-\xi ) \label{ex1:four}
\eea
For polarization 1 of Fig.~\ref{fig:pol}, equations \eqn{ex1:one}, 
\eqn{ex1:two} and \eqn{ex1:four} lead to
\bea
   r c_i ({\cal E}^Q_i - {\cal E}^Q_r) &=& \cos\almix c_t {\cal E}^{\tilde{Q}}_t \\
  -r c_i {\cal E}^{T_8}_r &=& \sin\almix c_t {\cal E}^{\tilde{Q}}_t \\
   \cos\almix ({\cal E}^Q_i + {\cal E}^Q_r) + \sin\almix {\cal E}^{T_8}_r &=&
      {\cal E}^{\tilde{Q}}_t
\eea
which can be solved for the amplitudes in Table \ref{tab:pol1-k_pos}, row 5.
Using \eqn{RT1} and \eqn{RT2} we obtain the reflection/transmission coefficients
of Table \ref{tab:RTcoeff-pol1}, row 5.

For polarization 2 of Fig.~\ref{fig:pol}, 
equations \eqn{ex1:one}, \eqn{ex1:two} 
and \eqn{ex1:three} lead to
\bea
   r ({\cal E}^Q_i + {\cal E}^Q_r) &=& \cos\almix {\cal E}^{\tilde{Q}}_t \\
   r {\cal E}^{T_8}_r &=& \sin\almix {\cal E}^{\tilde{Q}}_t \\
   c_i (\cos\almix ({\cal E}^Q_i - {\cal E}^Q_r) - \sin\almix {\cal E}^{T_8}_r)
      &=& c_t {\cal E}^{\tilde{Q}}_t 
\eea
which can similarly
be solved for the amplitudes in Table \ref{tab:pol2-k_pos}, row 5,
and reflection/transmission coefficients in Table \ref{tab:RTcoeff-pol2}, row 5.

\subsection{$T_8$ Higgsed, $X$ Higgsed}
When there is Higgsing in both regions,
the boundary condition equations \eqn{bc:D}-\eqn{bc:H} become
\begin{widetext}
\bea
   D^Q_{\perp}(\xi ) \Qvec - D^{\tilde{Q}}_{\perp}(-\xi ) \Qtvec &=& 
      \si^{T_8} \Tvec  + \si^X \Xvec \\
   E^Q_{\parallel}(\xi ) \Qvec - E^{\tilde{Q}}_{\parallel}(-\xi ) \Qtvec &=& 0 \\
   B^Q_{\perp}(\xi ) \Qvec - B^{\tilde{Q}}_{\perp}(-\xi ) \Qtvec &=& 0 \\
   H^Q_{\parallel}(\xi ) \Qvec - H^{\tilde{Q}}_{\parallel}(-\xi ) \Qtvec &=& 
      K^{T_8} \Tvec  + K^X \Xvec
\eea
\end{widetext}

In this case, we find
\bea
   E^Q_{\parallel}(\xi ) &=& E^{\tilde{Q}}_{\parallel}(-\xi ) = 0 \label{ex2:one}\\
   B^Q_{\perp}(\xi) &=& B^{\tilde{Q}}_{\perp}(-\xi ) = 0 \label{ex2:two}
\eea

For polarization 1 of Fig.~\ref{fig:pol}, either equation \eqn{ex2:one} or
\eqn{ex2:two} leads to the simple equations
\bea
   {\cal E}^Q_i + {\cal E}^Q_r &=& 0 \\
   {\cal E}^{\tilde{Q}}_t &=& 0
\eea
which shows that waves of this polarization are completely reflected with
a 180 degree phase shift.

For polarization 2 of Fig.~\ref{fig:pol}, either equation \eqn{ex2:one} or
\eqn{ex2:two} leads to the equally simple equations
\bea
   {\cal E}^Q_r - {\cal E}^Q_i &=& 0 \\
   {\cal E}^{\tilde{Q}}_t &=& 0
\eea
which shows that waves of this polarization are completely reflected with no
phase shift. 

\begin{table*}
\caption{Reflection and Transmission Coefficients for Polarization 1.
  For definitions see \eqn{angledefs}: $c_i$ and $c_t$ are the cosines
  of the incident and transmitted beams; $\almix$ is the mismatch between
  the generators of the free $U(1)$'s in the outside region ($Q$) and
  the inside region ($\Qtilde$); $r$ is a function of the permittivities
  and permeabilities of the two regions.
  \label{tab:RTcoeff-pol1}}
\begin{ruledtabular}
\newcommand{\shortstrut}{\rule[-1.5ex]{0em}{4ex}}
\newcommand{\tallstrut}{\rule[-2.6ex]{0em}{6.6ex}}
\begin{tabular}{llllll}
   \shortstrut Outer region ($T_8)$ & Inner region ($X$)
   & $R^Q$ & $R^{T_8}$ & $T^{\tilde{Q}}$ & $T^X$ \\
   \hline
   \shortstrut Higgsed & Higgsed & $1$ & 0 & 0 & 0 \\
   \hline
   \shortstrut Confined & Confined & $1$ & 0 & 0 & 0 \\
   \hline
   \tallstrut Free & Higgsed &
   $\dsp{\left(\frac{c_i \cos2\almix - r c_t}{c_i + r c_t}\right)^2}$ &
   $\dsp{\left(\frac{c_i \sin2\almix}{c_i + r c_t}\right)^2}$ &
   $\dsp{\frac{4 r c_i c_t \cos^2\! \almix}{(c_i + r c_t)^2}}$ & 0 \\
   \hline
   \tallstrut Higgsed & Free & 
   $\dsp{\left(\frac{c_i - r c_t}{c_i + rc_t}\right)^2}$ &
   0 &
   $\dsp{\frac{4 r c_i c_t \cos^2\! \almix}{(c_i + rc_t)^2}}$ &
   $\dsp{\frac{4 r c_i c_t \sin^2\! \almix}{(c_i + rc_t)^2}}$ \\
   \hline
   \tallstrut Free & Confined &
   $\dsp{\left(\frac{c_i - r c_t \cos 2\almix}{c_i + rc_t}\right)^2}$ &
   $\dsp{\left(\frac{r c_t \sin 2\almix}{c_i + rc_t}\right)^2}$ &
   $\dsp{\frac{4 r c_i c_t \cos^2\! \almix}{(c_i + rc_t)^2}}$ & 0 \\
   \hline
   \tallstrut Confined & Free &
   $\dsp{\left(\frac{c_i - r c_t}{c_i + r c_t}\right)^2}$ & 0 &
   $\dsp{\frac{4 r c_i c_t \cos^2\! \almix}{(c_i + r c_t)^2}}$ & 
   $\dsp{\frac{4 r c_i c_t \sin^2\! \almix}{(c_i + r c_t)^2}}$ \\
   \hline
   \tallstrut Higgsed & Confined &
   $\dsp{\left(\frac{c_i - r c_t \cos^2\! \almix}{c_i + r c_t \cos^2\! \almix}\right)^2}$ & 0 &
   $\dsp{\frac{4 r c_i c_t \cos^2\! \almix}{(c_i + r c_t \cos^2\! \almix)^2}}$ & 0 \\
   \hline
   \tallstrut Confined & Higgsed &
   $\dsp{\left(\frac{c_i \cos^2\! \almix - r c_t}{c_i \cos^2\! \almix + r c_t}\right)^2}$ & 0 &
   $\dsp{\frac{4 r c_i c_t \cos^2\! \almix}{(c_i \cos^2\! \almix + r c_t)^2}}$ & 0 \\
\end{tabular}
\end{ruledtabular}
\end{table*}
   
\begin{table*}
\caption{Reflection and Transmission Coefficients for Polarization 2
 \label{tab:RTcoeff-pol2}}
\begin{ruledtabular}
\newcommand{\shortstrut}{\rule[-1.5ex]{0em}{4ex}}
\newcommand{\tallstrut}{\rule[-2.6ex]{0em}{6.6ex}}
\begin{tabular}{llllll}
   \shortstrut Outer region ($T_8)$ & Inner region ($X$)
   & $R^Q$ & $R^{T_8}$ & $T^{\tilde{Q}}$ & $T^X$ \\
   \hline
   \shortstrut Higgsed & Higgsed & $1$ & 0 & 0 & 0 \\
   \hline
   \shortstrut Confined & Confined & $1$ & 0 & 0 & 0 \\
   \hline
   \tallstrut Free & Higgsed &
   $\dsp{\left(\frac{r c_i - c_t \cos2\almix}{r c_i + c_t}\right)^2}$ & 
   $\dsp{\left(\frac{c_t \sin2\almix}{r c_i + c_t}\right)^2}$ &
   $\dsp{\frac{4 r c_i c_t \cos^2\! \almix}{(r c_i + c_t)^2}}$ & 0 \\
   \hline
   \tallstrut Higgsed & Free &
   $\dsp{\left(\frac{r c_i - c_t}{r c_i + c_t}\right)^2}$ & 0 &
   $\dsp{\frac{4 r c_i c_t \cos^2\! \almix}{(r c_i + c_t)^2}}$ &
   $\dsp{\frac{4 r c_i c_t \sin^2\! \almix}{(r c_i + c_t)^2}}$ \\
   \hline
   \tallstrut Free & Confined &
   $\dsp{\left(\frac{r c_i \cos 2 \almix - c_t}{r c_i + c_t}\right)^2}$ &
   $\dsp{\left(\frac{r c_i \sin 2 \almix}{r c_i + c_t}\right)^2}$ &
   $\dsp{\frac{4 r c_i c_t \cos^2\! \almix}{(r c_i + c_t)^2}}$ & 0 \\
   \hline
   \tallstrut Confined & Free &
   $\dsp{\left(\frac{r c_i - c_t}{r c_i + c_t}\right)^2}$ & 0 &
   $\dsp{\frac{4 r c_i c_t \cos^2\! \almix}{(r c_i + c_t)^2}}$ &
   $\dsp{\frac{4 r c_i c_t \sin^2\! \almix}{(r c_i + c_t)^2}}$ \\
   \hline
   \tallstrut Higgsed & Confined &
   $\dsp{\left(\frac{r c_i \cos^2\! \almix - c_t}{rc_i \cos^2\! \almix + c_t}\right)^2}$ & 0 &
   $\dsp{\frac{4 r c_i c_t \cos^2\! \almix}{(r c_i \cos^2\! \almix + c_t)^2}}$ & 0 \\  
   \hline
   \tallstrut Confined & Higgsed &
   $\dsp{\left(\frac{r c_i - c_t \cos^2\!\almix}{r c_i + c_t \cos^2\!\almix}\right)^2}$ & 0 &
   $\dsp{\frac{4 r c_i c_t \cos^2\! \almix}{(r c_i + c_t \cos^2\!\almix)^2}}$ & 0 \\
\end{tabular}
\end{ruledtabular}
\end{table*}

\section{Summary and Discussion}
\label{sec:discussion}

We have studied reflection and transmission of gauge bosons
at the interface between
differently realized phases of a $U(1)\otimes U(1)$ gauge theory.
In order to allow gauge bosons to propagate, at least one linear
combination of the gauge generators must be free on each side of
the interface: this is taken to be $Q$ in the outer region ($z>0$)
and $\Qtilde$ in the inner region ($z<0$).
The other generator is $T_8$ in the outer region and $X$ in the
inner region. 
The possibilities for this other generator are that
it can be also free, Higgsed, or confined. 
The $(Q,T_8)$ basis may in general be rotated by an angle $\almix$
relative to the $(\Qtilde,X)$ basis.
Since the Free/Free boundary ($T_8$ free outside, $X$ free inside)
is trivial for any $\almix$, this means that there are
8 possible types of boundary.
The transmission and reflection coefficients for light arriving
at these different types of boundary are given in Tables
\ref{tab:RTcoeff-pol1} and \ref{tab:RTcoeff-pol2}. These are 
complicated so we give a qualitative summary in table \ref{tab:summary},
and we will now discuss the entries in that table.

\subsection{How the different boundaries behave}
\label{sec:summary}

\newcommand{\stack}[2]{ \begin{tabular}{l}#1 \\ #2\end{tabular} }
\begin{table*}
\caption{
Behavior of gauge bosons at an interface in $U(1)\otimes U(1)$ gauge theory
for various realizations of the gauge symmetries on each side.
The gauge bosons are assumed to arrive
as $Q$-photons from the ``outer'' phase.
\label{tab:summary}
}
\begin{tabular}{cc|ccc}
     &      & \multicolumn{3}{c}{{\bf Inner} ($z<0$), $\Qtilde$ free} \\[1ex]
     &      & $X$ Free & $X$ Higgsed & $X$ Confined \\
\hline
     & $T_8$ Free 
        & transmission 
        & \stack{$Q,T_8$ reflection}{$\Qtilde$ transmission}
        & \stack{$Q,T_8$ reflection}{$\Qtilde$ transmission} \\
\begin{tabular}{c} {\bf Outer}\\($z>0$)\\ $Q$ free \end{tabular}
     & $T_8$ Higgsed
        & \stack{$Q$ reflection}{$\Qtilde,X$ transmission} 
        & total reflection 
        & \stack{$Q$ reflection}{$\Qtilde$ transmission}   \\
     & $T_8$ Confined 
        & \stack{$Q$ reflection}{$\Qtilde,X$ transmission}
        & \stack{$Q$ reflection}{$\Qtilde$ transmission} & total reflection 
\end{tabular}
\end{table*}

If all generators everywhere are free
($T_8$ and $X$ both free, row 1 column 1 of table
\ref{tab:summary}), then there is no distinction between
the inner and outer regions other than a possible difference
in refractive index, so there will be transmission and reflection
as at a dielectric boundary like a glass-air boundary. 

If both generators in the outer region are free
($T_8$ free) but in the inner region $X$ is Higgsed or confined
(row 1 columns 2 and 3 of table \ref{tab:summary}), 
then the gauge bosons
are partially reflected and partially transmitted, depending on the
angle between $\Qtilde$ and $Q$. However, even though the incident
wave is pure $Q$ gauge bosons, there will be some additional $T_8$
bosons created and reflected back. The transmitted wave will
be pure $\Qtilde$ gauge bosons. 
Similarly, if both generators in the inner region are free
($X$ is free) and in the outer region $T_8$ is Higgsed or confined 
(column 1 rows 2 and 3 of table \ref{tab:summary}) then
there will be transmission of both $\Qtilde$ and $X$, adding up to make
a $Q$-photon.

If there is only one free generator in each region, $Q$ on the outside
and $\Qtilde$ on the inside, then the reflected wave must
be pure $Q$-photons and the transmitted wave must be pure $\Qtilde$ photons.
If the broken generator is Higgsed on one side and confined on the other
then there is partial reflection and partial transmission (row 2 column 3
and column 3 row 2 of table \ref{tab:summary}). This was the case
studied in \cite{manraj}.

If the broken generators on the inside and outside are both
Higgsed, or both confined then the behavior is very different.
Electromagnetic waves are {\em completely reflected} at an interface
between two Higgsed phases with different values of $\almix$ 
(row 2 column 2 of table \ref{tab:summary}) or between
two confined phases where the confined gauge fields are different
linear combinations of $Q$ and $T_8$
(row 3 column 3 of table \ref{tab:summary}). In addition, since one
polarization is flipped in each case while the other stays the same,
left circularly polarized waves are reflected as right circularly
polarized waves, and vice versa.
This raises an interesting puzzle: the Higgs/Higgs and confined/confined
boundaries both show total reflection independent of the value
of $\almix$. But when $\almix=0$ both phases have identical 
unbroken gauge generators, so the interface is just a boundary
between two media with different dielectric constants,
and there should be some transmission. In fact, in the limit 
$(\tilde\eps,\tilde\mu) \to (\eps,\mu)$ there
is no boundary, and there must be total transmission. This paradox
is analyzed below.

\subsection{Compatibility with previous results}

In Ref.~\cite{manraj}, Manuel and Rajagopal studied
the case where $X$ is Higgsed on the inside and
$T_8$ is confined on the outside, and our results for
that case agree with theirs. One of their main
conclusions was that it is possible to use light
reflection calculations to show that
there are magnetic monopoles in the QCD vacuum.
Their argument was that the situation they studied
corresponds to the boundary between the confining
QCD vacuum and color-superconducting quark matter, and
for that situation they derived the
confining boundary condition for $T_8$ color-magnetic flux,
which tells us there are $T_8$ magnetic monopoles in the boundary region, from
a few basic assumptions, namely:
(1) color is not Higgsed, so
  there are no color ($T_8$) supercurrents in the boundary layer;
(2) no gluons (ie $T_8$ gauge bosons) propagate in the confined phase;
(3) conservation of energy;
(4) Snell's law for the angles of reflection and transmission. 

This result can be obtained
more directly, 
without using light reflection calculations,
from considerations of static electromagnetic fields at an interface
using assumptions (1) and (2) alone.
Consider what must happen to $\Qtilde$ magnetic flux lines that arrive 
at the boundary from the quark matter side. Their $T_8$ component
cannot penetrate into the QCD vacuum region, since color is confined
there (assumption (2)), and they cannot be turned back into
the quark matter region by the Meissner effect because
there are no $T_8$ supercurrents in the boundary layer
(assumption (1)). So the flux lines have to end. This means that at the
edge of a color-confined phase there must be a boundary
layer of color magnetic monopoles that eat up any unwanted 
color magnetic flux that might try to enter the confined region.

\subsection{The singular $\almix\to 0$ limit}
\label{sec:singular}

We now turn to the paradoxical behavior of the Higgsed/Higgsed
and confined/confined interfaces, which seem to always reflect
all light even in the limit $\almix\to 0$, where the interface
becomes a typical dielectric boundary which ought to transmit
at least some light.  To understand
this we have to be careful about specifying the wavelength of the 
light that is incident on the boundary.

As mentioned in section \ref{sec:bc}, throughout
our calculations we have worked in the limit
of long wavelength relative to the penetration
depth, $\la \gg \xi$. This corresponds to the low frequency
limit, $\om \ll c/\xi$.
It turns out that, for the Higgsed/Higgsed
and confined/confined interfaces, the limit of low frequency does not commute
with the limit $\almix\to 0$ in which the unbroken $U(1)$'s
on either side of the boundary become the same.
An explicit calculation for the Higgs-Higgs boundary at 
finite $\almix$ and $\om$ 
is given in Appendix \ref{app:frequency}.

We can summarize the result as follows.
For polarization 1 (the argument for polarization 2 is analogous)
the transmission amplitude at low frequency ($\om\xi \ll c$)
and small $\almix$ is of the form
\beq
\frac{\om\xi}{\om\xi + i \almix^2 c}
\eeq
where $\xi$ is the penetration depth, and dimensionless
factors of order one (cosines of angles, etc) have been omitted.
In the limit where $\om\to 0$ first, $\om\xi \ll \almix^2 c \ll c$,
the transmission amplitude is zero: this is the total reflection 
expressed in the first two rows of
tables \ref{tab:RTcoeff-pol1} and \ref{tab:RTcoeff-pol2}.
In the limit where $\almix\to 0$ first,
$ \almix^2 c \ll\om\xi \ll c$, the transmission amplitude is
of order 1: this is what we expect when there is no mismatch
between the unbroken $U(1)$'s at the boundary.
We conclude that the paradox is resolved in this way:
at small $\almix$ there is total reflection
for frequencies  below $\almix^2 c/\xi$, but higher frequencies
are transmitted. As $\almix\to 0$ the range of reflected frequencies
becomes smaller and smaller, and finally disappears.

For most of the boundaries we studied, the two limits commute,
and we can, without ambiguity,
work at arbitrarily low frequency, and discuss how
the reflection and transmission depend on $\almix$. But for the
Higgs/Higgs and confined/confined boundaries the order of
the limits must be specified.

\subsection{Complementarity}
The complementarity principle \cite{complementarity} states that
for any Higgsed description of a gauge theory there should be
a corresponding confined description, so that there is
no way to distinguish a confined phase from a Higgs phase.
Since the Higgs phase involves condensation of electrically charged
fields, while the confined phase involves condensation
of magnetically charged fields, we expect that the
confined$\,\rightleftharpoons\,$Higgs mapping will involve
a magnetic$\,\rightleftharpoons\,$electric duality transformation.
Exchanging magnetic and electric fields converts polarization
1 into polarization 2 (see Fig.~\ref{fig:pol}), so the
confined$\,\rightleftharpoons\,$Higgs mapping will be

\beq
\label{duality}
\ba{rcl@{\qquad\quad}rcl}
   \vec{E} &\to& \vec{H}, & \vec{H} \to -\vec{E}, \\
   \vec{D} &\to& \vec{B}, & \vec{B} \to -\vec{D}, \\
   q_e &\to& q_m,         & q_m \to -q_e, \\
   \vec{J}_e &\to& \vec{J}_m, & \vec{J}_m \to -\vec{J}_e, \\
\multicolumn{6}{c}{ \tilde{\eps} \rightleftharpoons \tilde{\mu},} \\
\multicolumn{6}{c}{ r \rightleftharpoons 1/r } \\
\multicolumn{6}{c}{ \mbox{polarization 1} 
  \rightleftharpoons \mbox{polarization 2} } \\
\multicolumn{6}{c}{ \mbox{Higgsed} \rightleftharpoons \mbox{Confined} }\\
\ea
\eeq

Since the reflection and transmission coefficients are 
related to the energy and momentum flow in the scattering process,
they are directly observable, and should be invariant under the
duality transformation \eqn{duality}.
Inspecting tables \ref{tab:RTcoeff-pol1} and
\ref{tab:RTcoeff-pol2} we see that this is indeed the case.  For
example, the reflection and transmission coefficients for the
Higgsed-Free boundary (third line of table \ref{tab:RTcoeff-pol1}) are
transformed into those for the Confined-Free boundary (fifth line in
table \ref{tab:RTcoeff-pol2}).
In other words, if we shine light on a boundary and obtain the
results of table \ref{tab:RTcoeff-pol1} line 3, then we could
not distinguish whether the outside phase is Higgsed or confined.

This means that by measuring only the reflected and transmitted
indensities, we can only distinguish 4 of the 8 types of non-trivial
boundary. What is clear from tables \ref{tab:RTcoeff-pol1} and
\ref{tab:RTcoeff-pol2} is that this ambiguity only exists
as a single global choice. There is not a separate confined vs.~Higgs
choice for each phase independently. This is exactly what we
expect from the principle of complementarity.

One might naively think that it should be possible to overcome
this ambiguity by measuring the electric and magnetic fields 
(which are also gauge-invariant and physically measureable)
directly. Carefully constructing the corresponding thought-experiment
shows that this does not in fact overcome the ambiguity: we discuss
this in appendix \ref{app:fields}.

\subsection{Future directions}
As mentioned in the introduction, the $U(1)\otimes U(1)$ system
arises in various contexts within particle physics, and the
results of this paper may be applied to domain walls
or phase boundaries in those contexts. The same
formalism can also be used for more general
gauge groups, as in the work of
Manuel and Rajagopal \cite{manraj}.
Quark matter provides a possible area of application, since it has a rich
phase diagram, including a variety of patterns of 
confinement or Higgsing of various subgroups of the
$SU(3)_{\rm color}\otimes U(1)_Q$ gauge group \cite{Reviews}.

Finally, in our analysis we only concerned ourselves with the gauge
symmetries, not with any global symmetries.  If massless fermionic
fields are included in the theory then chiral symmetry complicates the
complementarity principle \cite{complementarity_with_fermions}.  It
would be interesting to see how this affects the distinguishability of
our $U(1)\otimes U(1)$ interfaces.  One immediate question is the
contradiction between Ref.~\cite{complementarity_with_fermions},
which predicts that chiral
symmetries will not not be broken in weakly-coupled Higgsed
phases, and the accepted picture of high-density quark
matter, according to which CFL pairing produces Higgs breaking of the
color gauge symmetry and simultaneously breaks chiral symmetry.
This is crucial to the concept of quark-hadron continuity, which
identifies the CFL phase as a controlled continuation of the
confined phase.

\begin{acknowledgments}
We thank Krishna Rajagopal for helpful
clarifications of the results of Ref.~\cite{manraj}. 
This work is supported in part by the U.S. Department of Energy
under grant number DE-FG02-91ER40628.
\end{acknowledgments}

\appendix
\section{Non-zero-frequency effects}
\label{app:frequency}

The macroscopic calculations of section \ref{sec:bc} are performed
under the simplifying assumption that the frequencies are very small,
and therefore the time-derivative terms in the Maxwell equations are
neglected. It is also assumed that the Higgsed or confined fields are
quickly screened, so the field amplitudes are set to zero from the
beginning. The advantage of this approach is that the spatial behavior
of the screened fields and the screening currents does not have to be
determined, so the solution is straightforward. However, any finite
frequency effects are thrown away, and as mentioned in section
\ref{sec:singular}, Higgsed/Higgsed and confined/confined interfaces
have singular behavior in the $\almix \to 0$ limit.  To rectify this
problem, we performed the calculation again for the Higgsed/Higgsed
combination, keeping the contributions of screened fields and finite
frequency.

First, we briefly review the behavior of electromagnetic fields in a
superconductor.  In addition to the Maxwell equations
(\ref{eqn:maxwell}), we have the London equations
(see Ref.~\cite{ashcroftmermin}, chapter 34)
\bea
   \label{eqn:london}
   \frac{d \vec{J}}{d t} &=& \gamma \vec{E}, \notag \\
   \vec{\nabla} \times \vec{J} &=& -\gamma \vec{B}
\eea
that describe how the supercurrents respond to applied fields. The parameter 
$\gamma$ depends on microscopic details such as the density and charge of
Cooper pairs that make up the supercurrents, but the details are not
important for this discussion. Inserting equations (\ref{eqn:london})
into the ``curl'' equation for $\vec{B}$, we obtain the wave equation
for the magnetic field in the superconductor,
\beq
  \nabla^2 \vec{B} = \mu \gamma \vec{B} + \frac{1}{c^2} \frac{\p^2 \vec{B}}{\p t^2}
\eeq
From dimensional considerations the definition of the screening length $\xi$ is defined
as
\beq
  \xi \equiv \frac{1}{\sqrt{\mu \gamma}}
\eeq
and the solutions of the wave equation have the form
\beq
  \vec{B} = \vec{B_0} \exp \left[ \vec{\kappa} \cdot \vec{x} - i \om t \right].
\eeq
Plugging this solution back into the wave equation obtains the magnitude of the
wavevector, 
\beq
  |\kappa | = \pm \sqrt{\frac{1}{\xi^2} - \frac{\om^2}{c^2}}.
\eeq
For frequencies less than $c/\xi$, the waves are completely damped,
while for frequencies greater than $c/\xi$, the waves propagate
without any damping. For $\om = c/\xi$, the wave has no spatial
variation and only oscillates in time.  The choice of the
positive or negative solution for the wavevector depends on the
boundary conditions of the superconducting phase. We can obtain identical 
wave equations for $\vec{J}$ and
$\vec{E}$; since we are still interested in the low-frequency limit,
we can use the limiting value $|\kappa | = \pm \xi^{-1}$ to obtain
the magnitudes of the current and the electric field as
\bea
  |J_0| = \mp \xi \gamma |B_0|, \notag \\
  |E_0| = \pm i \om \xi |B_0|
\eea

For the Higgsed/Higgsed phase combination, the Higgsed fields on either side
of the boundary will satisfy the equations above. Explicitly, we have
\bea
\vec{E}^Q_i &=& {\cal{E}}^Q_i \vec n_i 
  \exp(i(\vec{k}_i \cdot \vec{x}-\om t))\ , \notag \\
\vec{E}^Q_r &=& {\cal{E}}^Q_r \vec n_r
  \exp(i(\vec{k}_r \cdot \vec{x}-\om t))\ , \notag \\
\vec{E}^{T_8}_r &=& {\cal{E}}^{T_8}_r \vec n_r
  \exp(-\vec{\kappa}_r \cdot \vec{x}-i\om t)\ , \notag \\
\vec{E}^{\tilde{Q}}_t &=& {\cal{E}}^{\tilde{Q}}_t \vec{n}_t 
  \exp(i(\vec{k}_t \cdot \vec{x}-\om t)) , \notag \\ 
\vec{E}^X_t &=& {\cal{E}}^X_t \vec{n}_t 
  \exp(+\vec{\kappa}_t \cdot \vec{x}-i\om t))\ , 
\eea
The ``${\cal{E}}$'' amplitudes are the magnitudes of the electric field at
the boundary itself ($z=0$); all screening is due to the spatial terms.

Now we will rewrite the boundary condition equations keeping
everything that was thrown away previously. The ``curl'' equations are
sufficient to solve for the field amplitudes.  We obtain
\bea
  && E^Q_\parallel \Qvec + E^{T_8}_\parallel \Tvec - E^\Qtilde_\parallel \Qtvec
     - E^X_\parallel \Xvec \notag\\
&& = \frac{i\om}{\kappa_z} B^{T_8}_\parallel \Tvec 
     + \frac{i\om}{\tilde{\kappa}_z} B^X_\parallel \Xvec \\[2ex]
&&  B^Q_\parallel \Qvec + \left( 1 - \frac{1}{\xi \kappa_z} \right) B^{T_8}_\parallel \Tvec
     - B^\Qtilde_\parallel \Qtvec \notag \\
 &&  \makebox[3em]{~}
  - \left( 1 + \frac{1}{\tilde{\xi} \tilde{\kappa}_z} \right) 
    B^X_\parallel \Xvec \notag \\
 &&  = - \frac{i\om}{c^2 \kappa_z} E^{T_8}_\parallel \Tvec
     - \frac{i\om}{\tilde{c}^2 \tilde{\kappa}_z} E^X_\parallel \Xvec 
\eea

\newcommand{\nfrac}{\frac{\tilde{n}}{n}}

For polarization 1 of Figure \ref{fig:pol}, 
the solutions for the amplitudes are
\beq
\ba{rcl}
   \dsp\frac{{\cal E}^Q_r}{{\cal E}^Q_i} &=&\dsp
     \frac{\om \tilde{\xi} (c_i - \cos^2\! \almix\nfrac c_t) + i c (1 + c_t)\sin^2\! \almix }
          {\om \tilde{\xi} (c_i + \cos^2\! \almix\nfrac c_t) - i c (1 + c_t)\sin^2\! \almix }
   \\[3ex]
   \dsp\frac{{\cal E}^{T_8}_r}{{\cal E}^Q_i} &=&\dsp
     \frac{\sin 2\almix \om \xi c_i (1 + c_t)}
          {\om \tilde{\xi} (c_i + \cos^2\! \almix\nfrac c_t) - i c (1 + c_t)\sin^2\! \almix }
   \\[3ex]
   \dsp\frac{{\cal E}^\Qtilde_t}{{\cal E}^Q_i} &=&\dsp
     \frac{2 \cos \almix \om \tilde{\xi} c_i }
          {\om \tilde{\xi} (c_i + \cos^2\! \almix\nfrac c_t) - i c (1 + c_t)\sin^2\!\almix }
   \\[3ex]
   \dsp\frac{{\cal E}^X_t}{{\cal E}^Q_i} &=&\dsp
     \frac{2 \sin \almix \om \tilde{\xi} c_i }
          {\om \tilde{\xi} (c_i + \cos^2\! \almix\nfrac c_t) - i c (1 + c_t)\sin^2\! \almix }
   .
\ea
\eeq

Taking the $\om \to 0$ limit, we recover the amplitudes of the first row of 
Table \ref{tab:pol1-k_pos} presented below in Appendix \ref{app:fields}. However,
more importantly, taking the $\almix \to 0$ limit first, we obtain
\bea
{\cal E}^Q_r &=& {\cal E}^Q_i \left[ \frac{c_i - \nfrac c_t}{c_i + \nfrac c_t} \right] \notag \\
{\cal E}^{T_8}_r &\to& 0 \notag \\
{\cal E}^\Qtilde_t &=& {\cal E}^Q_i \left[ \frac{2 c_i}{c_i + \nfrac c_t} \right] \notag \\
{\cal E}^X_t &\to& 0,  
\eea
which are the normal reflection and refraction amplitudes from electrodynamics.
Similarly, for polarization 2 of Figure \ref{fig:pol}, the solutions for the amplitudes
are
\begin{widetext}
\beq
\ba{rcl} 
   \dsp\frac{{\cal E}^Q_r}{{\cal E}^Q_i} &=& \dsp
     \frac{2 \om \left[ \xi (1+c_i^2)(1+c_t)(\nfrac c_i - \cos^2\! \almix c_t) - 
                        \tilde{\xi} (1+c_i)(1+c_t^2)(\cos^2\! \almix \nfrac c_i - c_t) \right]
         + 2i \sin^2\! \almix c c_i c_t (1+c_i)(1+c_t)}
          {2 \om \left[ \xi (1+c_i^2)(1+c_t)(\nfrac c_i + \cos^2\! \almix c_t) - 
                        \tilde{\xi} (1+c_i)(1+c_t^2)(\cos^2\! \almix \nfrac c_i + c_t) \right]
         + 2i \sin^2\! \almix c c_i c_t (1+c_i)(1+c_t) }
   \\[3ex]
   \dsp\frac{{\cal E}^{T_8}_r}{{\cal E}^Q_i}&=& \dsp
     \frac{- 2 \om \xi \sin 2\almix c_i^2 c_t (1+c_t)}
          {2 \om \left[ \xi (1+c_i^2)(1+c_t)(\nfrac c_i + \cos^2\! \almix c_t) - 
                        \tilde{\xi} (1+c_i)(1+c_t^2)(\cos^2\! \almix \nfrac c_i + c_t) \right]
         + 2i \sin^2\! \almix c c_i c_t (1+c_i)(1+c_t)  }
   \\[3ex]
   \dsp\frac{{\cal E}^\Qtilde_t}{{\cal E}^Q_i}&=& \dsp
     \frac{4 \om \cos \almix c_i \left[\xi (1+c_i^2)(1+c_t) - \tilde{\xi} (1+c_i)(1+c_t^2)\right]}
          {2 \om \left[ \xi (1+c_i^2)(1+c_t)(\nfrac c_i + \cos^2\! \almix c_t) - 
                        \tilde{\xi} (1+c_i)(1+c_t^2)(\cos^2\! \almix \nfrac c_i + c_t) \right]
         + 2i \sin^2\! \almix c c_i c_t (1+c_i)(1+c_t)  }
   \\[3ex]
   \dsp\frac{{\cal E}^X_t}{{\cal E}^Q_i}&=& \dsp
     \frac{4 \om \tilde{\xi} \sin \almix c_i (1+c_i) c_t^2}
          {2 \om \left[ \xi (1+c_i^2)(1+c_t)(\nfrac c_i + \cos^2\! \almix c_t) - 
                        \tilde{\xi} (1+c_i)(1+c_t^2)(\cos^2\! \almix \nfrac c_i + c_t) \right]
         + 2i \sin^2\! \almix c c_i c_t (1+c_i)(1+c_t)  }
   .
\ea
\eeq
\end{widetext}
Once again, taking the $\om \to 0$ limit, we recover the amplitudes of the first row of Table 
\ref{tab:pol2-k_pos} presented below in Appendix \ref{app:fields}. Taking the $\almix \to 0$ limit first,
we obtain
\bea
{\cal E}^Q_r &=& {\cal E}^Q_i \left[ \frac{\nfrac c_i - c_t}{\nfrac c_i + c_t} \right] \notag \\
{\cal E}^{T_8}_r &\to& 0 \notag \\
{\cal E}^\Qtilde_t &=& {\cal E}^Q_i \left[ \frac{2 c_i}{\nfrac c_i + c_t} \right] \notag \\
{\cal E}^X_t &\to& 0,  
\eea
the normal reflection and refraction amplitudes for perpedicularly
polarized light.

This shows that our ``singular'' limit problem is actually an
order-of-limits problem. For most of the possible phase combinations,
we could take the $\om \to 0$ limit at the beginning and not encounter
any problems, but for the Higgsed/Higgsed or confined/confined phases,
that is incorrect. Although the frequency drops out in the $\almix \to
0$ limit, we need to keep a nonzero frequency value to obtain the
correct expression.

\section{Field strengths and complementarity}
\label{app:fields}

In tables \ref{tab:pol1-k_pos} and \ref{tab:pol2-k_pos} we show the
reflection and transmission {\em amplitudes}, i.e. the ratios between
electric field strengths in the incident, reflected, and transmitted beams.
It is clear that the transmission amplitudes
do not show invariance under the duality
transformation \eqn{duality}. Does this mean that measurements
of electric and magnetic fields can overcome the complementarity
ambiguity and distinguish a Higgsed phase from a confined phase?
In this appendix, we show that although electric and magnetic fields
are gauge-invariant quantities, what can actually be measured is
the force exerted on a charge, so that even experiments that seem
to directly measure field strengths suffer from the 
Higgsed/confined ambiguity.

For illustrative purposes, we calculate the Lorentz force on a test
charge in the inner phase due to electromagnetic waves transmitted
from the outer phase. First, we will calculate the force in 
the case where the outer phase is confined and the inner phase is Higgsed; then
we will calculate the force in the dual picture, where the outer phase
is Higgsed and the inner phase is confined. 
We will see that although the transmission amplitudes are not
invariant under the duality transformation, the physically
measureable quantity, force, is invariant.

\begin{table*}
\caption{Reflected and Transmitted Amplitudes for Polarization 1
  \label{tab:pol1-k_pos}}
\begin{ruledtabular}
\newcommand{\shortstrut}{\rule[-1.5ex]{0em}{4ex}}
\newcommand{\tallstrut}{\rule[-2.6ex]{0em}{6.6ex}}
\begin{tabular}{llllll}
   \shortstrut Outer region ($T_8)$ & Inner region ($X$)
   & ${\cal E}^Q_r / {\cal E}^Q_i$ & ${\cal E}^{T_8}_r / {\cal E}^Q_i$
   & ${\cal E}^{\tilde{Q}}_t / {\cal E}^Q_i$ & ${\cal E}^X_t / {\cal E}^Q_i$ \\
   \hline
   \shortstrut Higgsed & Higgsed & $-1$ & 0 & 0 & 0 \\
   \hline
   \shortstrut Confined & Confined & $1$ & 0 & 0 & 0 \\
   \hline
   \tallstrut Free & Higgsed &
   $\dsp{\frac{c_i \cos2\almix - r c_t}{c_i + r c_t}}$ &
   $\dsp{\frac{c_i \sin2\almix}{c_i + r c_t}}$ &
   $\dsp{\frac{2 c_i \cos\almix}{c_i + r c_t}}$ & 0 \\
   \hline
   \tallstrut Higgsed & Free & 
   $\dsp{\frac{c_i - r c_t}{c_i + rc_t}}$ &
   0 &
   $\dsp{\frac{2 c_i \cos \almix}{c_i + rc_t}}$ &
   $\dsp{\frac{-2 c_i \sin \almix}{c_i + rc_t}}$ \\
   \hline
   \tallstrut Free & Confined &
   $\dsp{\frac{c_i - r c_t \cos 2\almix}{c_i + rc_t}}$ &
   $\dsp{\frac{-r c_t \sin 2\almix}{c_i + rc_t}}$ &
   $\dsp{\frac{2 c_i \cos \almix}{c_i + rc_t}}$ & 0 \\
   \hline
   \tallstrut Confined & Free &
   $\dsp{\frac{c_i - r c_t}{c_i + r c_t}}$ & 0 &
   $\dsp{\frac{2 c_i \cos\almix}{c_i + r c_t}}$ & 
   $\dsp{\frac{-2 c_i \sin\almix}{c_i + r c_t}}$ \\
   \hline
   \tallstrut Higgsed & Confined &
   $\dsp{\frac{c_i - r c_t \cos^2\! \almix}{c_i + r c_t \cos^2\! \almix}}$ & 0 &
   $\dsp{\frac{2 c_i \cos \almix}{c_i + r c_t \cos^2\! \almix}}$ & 0 \\
   \hline
   \tallstrut Confined & Higgsed &
   $\dsp{\frac{c_i \cos^2\! \almix - r c_t}{c_i \cos^2\! \almix + r c_t}}$ & 0 &
   $\dsp{\frac{2 c_i \cos\almix}{c_i \cos^2\! \almix + r c_t}}$ & 0 \\
\end{tabular}
\end{ruledtabular}
\end{table*}
   
\begin{table*}
\caption{Reflected and Transmitted Amplitudes for Polarization 2
 \label{tab:pol2-k_pos}}
\begin{ruledtabular}
\newcommand{\shortstrut}{\rule[-1.5ex]{0em}{4ex}}
\newcommand{\tallstrut}{\rule[-2.6ex]{0em}{6.6ex}}
\begin{tabular}{llllll}
   \shortstrut Outer region ($T_8)$ & Inner region ($X$)
   & ${\cal E}^Q_r / {\cal E}^Q_i$ & ${\cal E}^{T_8}_r / {\cal E}^Q_i$ 
   & ${\cal E}^{\tilde{Q}}_t / {\cal E}^Q_i$ & ${\cal E}^X_t / {\cal E}^Q_i$ \\
   \hline
   \shortstrut Higgsed & Higgsed & $1$ & 0 & 0 & 0 \\
   \hline
   \shortstrut Confined & Confined & $-1$ & 0 & 0 & 0 \\
   \hline
   \tallstrut Free & Higgsed &
   $\dsp{\frac{r c_i - c_t \cos2\almix}{r c_i + c_t}}$ & 
   $\dsp{\frac{-c_t \sin2\almix}{r c_i + c_t}}$ &
   $\dsp{\frac{2 c_i \cos\almix}{r c_i + c_t}}$ & 0 \\
   \hline
   \tallstrut Higgsed & Free &
   $\dsp{\frac{r c_i - c_t}{r c_i + c_t}}$ & 0 &
   $\dsp{\frac{2 c_i \cos \almix}{r c_i + c_t}}$ &
   $\dsp{\frac{-2 c_i \sin \almix}{r c_i + c_t}}$ \\
   \hline
   \tallstrut Free & Confined &
   $\dsp{\frac{r c_i \cos 2 \almix - c_t}{r c_i + c_t}}$ &
   $\dsp{\frac{r c_i \sin 2 \almix}{r c_i + c_t}}$ &
   $\dsp{\frac{2 c_i \cos \almix}{r c_i + c_t}}$ & 0 \\
   \hline
   \tallstrut Confined & Free &
   $\dsp{\frac{r c_i - c_t}{r c_i + c_t}}$ & 0 &
   $\dsp{\frac{2 c_i \cos\almix}{r c_i + c_t}}$ &
   $\dsp{\frac{-2 c_i \sin\almix}{r c_i + c_t}}$ \\
   \hline
   \tallstrut Higgsed & Confined &
   $\dsp{\frac{r c_i \cos^2\! \almix - c_t}{rc_i \cos^2\! \almix + c_t}}$ & 0 &
   $\dsp{\frac{2 c_i \cos \almix}{r c_i \cos^2\! \almix + c_t}}$ & 0 \\  
   \hline
   \tallstrut Confined & Higgsed &
   $\dsp{\frac{r c_i - c_t \cos^2\!\almix}{r c_i + c_t \cos^2\!\almix}}$ & 0 &
   $\dsp{\frac{2 c_i \cos\almix}{r c_i + c_t \cos^2\!\almix}}$ & 0 \\
\end{tabular}
\end{ruledtabular}
\end{table*}

A linearly polarized electromagnetic wave is sent from the outside,
through the interface, to the inside, where its effect on a test charge
is measured. On the outside, we calibrate the
wave by measuring how it causes an electric
$Q$ charge to move, and from the induced motion we measure the 
electric field strength ${\cal E}_i$.
On the inside, the transmitted wave causes an electric $\Qtilde$ charge to move, 
and the resulting motion allows calculation of the force. For our example,
we assume the wave to be in polarization 1 of figure \ref{fig:pol}.
As we have calculated in this paper, the transmitted $\Qtilde$-fields and
the force are
\bea
\label{pictureoneforce}
  &\vec{E}_t& = -\hat{z} {\cal{E}}_t, \notag \\
  &\vec{B}_t& = (-c_t \hat{x} + s_t \hat{y}) 
    \frac{1}{\tilde{c}} {\cal{E}}_t, \notag \\
  &\vec{F}& = -q_e {\cal{E}}_i 
      \left( \frac{2 c_i \cos 2\almix}{c_i + r c_t} \right) \times \notag \\
      &~&\left( \hat{z} \left( 1 + \frac{v_x s_t + v_y c_t}{\tilde{c}} \right)
             +\hat{y} \frac{v_z c_t}{\tilde{c}} +\hat{x} \frac{v_z s_t}{\tilde{c}} \right)
\eea

Now transform to the dual picture, where the outer phase
is Higgsed and the inner phase is confined, using the transformation
\eqn{duality}. Our calibration experiment now appears to have involved
a magnetic charge, feeling a ``Lorentz'' force 
\beq
   \vec{F} = q_m (\vec{H} - \vec{v} \times \vec{D}).
\eeq
The transmitted wave is in polarization state 2, with
\bea
   &\vec{H}& = -\hat{z} \sqrt{\frac{\tilde{\eps}}{\tilde{\mu}}} {\cal{E'}}_t, \notag \\
   &\vec{D}& = (c_t \hat{x} - s_t \hat{y}) \tilde{\eps} {\cal{E'}}_t, \notag \\
   &\vec{F}& = -q_m {\cal{E'}}_i \sqrt{\frac{\tilde{\eps}}{\tilde{\mu}}} 
      \left( \frac{2 c_i \cos 2\almix}{r c_i + c_t} \right) \times \notag \\
      &~&\left( \hat{z} \left( 1 + \frac{v_x s_t + v_y c_t}{\tilde{c}} \right)
             +\hat{y} \frac{v_z c_t}{\tilde{c}} +\hat{x} \frac{v_z s_t}{\tilde{c}} \right)
\eea
However, ${\cal{E'}}_i$ and ${\cal{E}}_i$ are not equal; because of the switch between
electric and magnetic fields, the amplitude of the waves at their source will be calibrated
so that 
\beq
   {\cal{E'}}_i = \sqrt{\frac{\mu}{\eps}} {\cal{E}}_i.
\eeq
Finally, we find that, in terms of the original incident amplitude ${\cal{E}}_i$, the
force measured in the inner phase is
\bea
\label{picturetwoforce}
  &\vec{F}& = -q_m {\cal{E}}_i r
      \left( \frac{2 c_i \cos 2\almix}{r c_i + c_t} \right) \times \notag \\
      &~&\left( \hat{z} \left( 1 + \frac{v_x s_t + v_y c_t}{\tilde{c}} \right)
             +\hat{y} \frac{v_z c_t}{\tilde{c}} +\hat{x} \frac{v_z s_t}{\tilde{c}} \right)
\eea

By taking the force calculated in the first picture (equation \eqn{pictureoneforce}), 
applying the duality transformation $r \to 1/r$ and then replacing
$q_e$ by $q_m$ (which were assumed to have equal magnitudes), we end up with the
expression of the force measured in the second picture (equation \eqn{picturetwoforce}).  
Since the two pictures are equivalent, the ambiguity remains and cannot be resolved by an 
attempt to measure the field amplitudes.

\end{document}